\documentclass[aps,prb,twocolumn,superscriptaddress,showpacs,floatfix]{revtex4}

\usepackage{times}
\usepackage{amsmath,bm,amsfonts}
\usepackage{dcolumn}
\usepackage{graphicx}
\usepackage{latexsym}

\usepackage{color}

\begin{document}

\title{Mott Insulating Ground State and its Proximity to Spin-Orbit
  Insulators in Na$_{2}$IrO$_{3}$}

\author{Hosub \surname{Jin}}
\author{Heungsik \surname{Kim}}
\affiliation{Department of Physics and Astronomy, CSCMR, Center for Theoretical Physics,
    Seoul National University, Seoul 151-747, Korea}
\author{Hogyun \surname{Jeong}}
\affiliation{Computational Science and Technology Interdisciplinary
  Program, Seoul National University, Seoul 151-747, Korea}
\author{Choong H. \surname{Kim}}
\affiliation{Department of Physics and Astronomy, CSCMR, Center for Theoretical Physics,
    Seoul National University, Seoul 151-747, Korea}
\author{Jaejun \surname{Yu}}
\email[Corresponding author. Electronic address:]{jyu@snu.ac.kr}
\affiliation{Department of Physics and Astronomy, CSCMR, Center for Theoretical Physics,
    Seoul National University, Seoul 151-747, Korea}

\date{\today }

\begin{abstract}
  We present an anti-ferromagnetically ordered ground state of
  Na$_{2}$IrO$_{3}$ based on density-functional-theory calculations
  including both spin-orbit coupling and on-site Coulomb interaction
  $U$. We show that the splitting of $e_{g}'$ doublet states by the strong
  spin-orbit coupling is mainly responsible for the intriguing nature of
  its insulating gap and magnetic ground state. Due to its proximity
  to the spin-orbit insulator phase, the magnetic ordering as obtained
  with finite $U$ is found to exhibit a strong in-plane anisotropy. The
  phase diagram of Na$_{2}$IrO$_{3}$ suggests a possible interplay between
  spin-orbit insulator and Mott anti-ferromagnetic insulator phases.
\end{abstract}


\maketitle

Recently, the role of spin-orbit coupling (SOC) has attracted great
attention in many fields of condensed matter physics. In multiferroic
materials, for example, SOC combined with a large electron-lattice
interaction has been suggested to be responsible for the multiferroic
behavior which exhibit both non-collinear magnetic ordering and lattice
polarization \cite{Kimura03,Hur04}. SOC is also indispensable to anomalous
Hall and spin Hall effects where Hall and spin Hall currents are generated
by an external electric field, respectively \cite{PhysRev.95.1154,
  ShuichiMurakami09052003,PhysRevLett.92.126603}.  In particular, the
quantum spin Hall effect has led to the notion of topological insulators,
new states of quantum matter \cite{zhang09,xia09}. While they have bulk
energy gaps generated by the SOC, topological insulators are characterized
by the presence of gapless surface states which are protected by
time-reversal symmetry \cite{kane:146802}.

Another manifestation of strong SOC combined with on-site Coulomb
interactions is the $j_{\mathrm{eff}}$=1/2 Mott insulator discovered in
Sr$_{2}$IrO$_{4}$, one of the 5$d$ transition-metal oxides
\cite{kim:076402, moon:226402}. The novel spin-orbit integrated state with
$j_{\mathrm{eff}}$=1/2 arises from the combined action of both strong SOC
and intermediate on-site Coulomb interactions within the Ir 5$d$ $t_{2g}$
manifold.  In addition, there has been a theoretical proposal on the room
temperature quantum spin Hall effect in Na$_2$IrO$_3$ based on the
$j_{\mathrm{eff}}$=1/2 physics \cite{shitade-2008}, where the honeycomb
lattice consisting of edge-shared IrO$_6$ octahedra in each Ir-O layer was
considered to be an ideal realization of the Kane-Mele model, where
hopping integrals between the $j_{\mathrm{eff}}$=1/2 states at the Fermi
level was assumed to be an essential ingredient for the quantum spin Hall
effect \cite{ kane:146802,PhysRevLett.95.226801}. Since the crystal
structure and local environment of Ir atoms in Na$_2$IrO$_3$ are different
from those of Sr$_{2}$IrO$_{4}$, however, it is necessary to clarify the
electronic and magnetic structures of the Ir 5$d$ manifold in this
Na$_2$IrO$_3$ compound with hexagonal lattice.

In this paper, we present novel electronic structure and magnetic
properties of Na$_2$IrO$_3$ by carrying out density-functional-theory
(DFT) calculations including both SOC $\lambda_{\mathrm{SO}}$ and on-site
Coulomb interaction.  We observe that a new form of spin-orbit coupled
states emerges from the $e_{g}'$ doublet states near the Fermi level
($E_{\mathrm{F}}$) and determines the intriguing nature of its insulating
gap. With an effective on-site Coulomb interaction parameter
$U=2.0$ eV, the ground state of Na$_2$IrO$_3$ is
found to be an antiferromagnetic (AFM) insulator with the ordered moments
lying down within the honeycomb lattice of Ir atoms. The large splitting
of the $e_{g}'$ doublet by the strong SOC is related to the strong
in-plane anisotropy of magnetic ordering. Considering the role of SOC, we
propose a phase diagram in the $\lambda_{\textrm{SO}}$--$U$ parameter
space which features a phase boundary between AFM Mott insulators and SO
insulators. By estimating the exchange couplings between
neighboring Ir atoms, we suggest a possible frustration of magnetic
ordering in its ground state, which is consistent with a recent
  experiment \cite{Takagi}.

In order to examine the effects of both SOC and on-site Coulomb
interaction on the electronic structure of Na$_2$IrO$_3$, it is necessary
to treat both SOC and $U$ on an equal footing in the description of Ir
5$d$ states.  To identify the role of each contribution as well as the
interplay between them, we carried out DFT calculations within the
local-density approximation (LDA), LDA including SOC (LDA+SO), and LDA+$U$
including SOC (LDA+$U$+SO) respectively. For the calculations, we used the
DFT code, OpenMX \cite{openmx}, based on the
linear-combination-of-pseudo-atomic-orbitals method
\cite{PhysRevB.67.155108}, where both the LDA+$U$ method \cite{han:045110}
and the SOC contribution were included via a relativistic $j$-dependent
pseudo-potential scheme in the non-collinear DFT formalism.  Double
valence and single polarization orbitals were used as basis sets, which
were generated by a confinement potential scheme with cutoff radii of 7.0,
7.0 and 5.0 a.u. for Na, Ir, and O atoms respectively.  We used a
(14$\times$14$\times$14) \textbf{k}-point grid for the k-space
integration.

Up to our knowledge there is no crystal structure data for Na$_2$IrO$_3$
published yet. Thanks to the preliminary information provided by
Takagi\cite{Takagi}, we were able to construct a minimal unit-cell
containing two formula units based on the hexagonal structure of
Na$_{2}$RuO$_{3}$ \cite{Kailash04}, a sibling compound of Na$_2$IrO$_3$.
The crystal structure of Na$_2$IrO$_3$ can be viewed as an alternate
stacking of (Ir$_{2/3}$Na$_{1/3}$)O$_2$ and Na layers. Edge-shared IrO$_6$
octahedra form a honeycomb lattice of Ir atoms.  Na atoms are placed at
the center of each hexagon. Upper and lower triangle oxygens are rotated
by 3.5$^{\circ}$ to shorten the Ir-O distance.  The positions of atoms in
the unit cell were determined through the full structural optimization by
the LDA calculations with 0.5$\times 10^{-3}$ Hatree/{\AA} of force
criterion. There is a possible stacking disorder in the types of the
Na-layers relative to the (Ir$_{2/3}$Na$_{1/3}$)O$_2$ layers. We have
checked the effect of different stacking sequences and observed a
negligible change in the energy dispersions. Since the basic electronic
structure is dominated by the in-plane Ir-O hybridization and remains
intact regardless of the stacking sequence, we will focus on the
electronic structure without structural disorder hereafter.

\begin{figure}
 \centering\includegraphics[width=8cm]{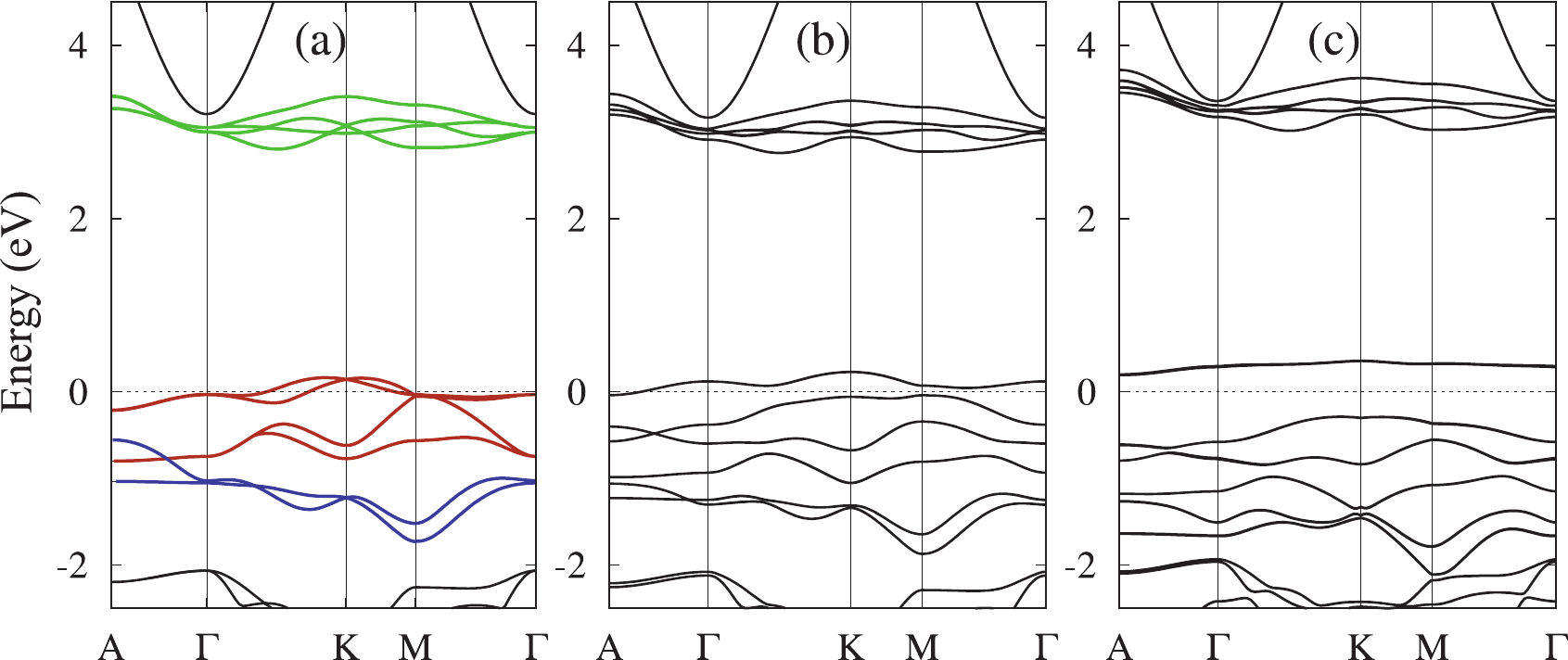}
 \caption{(Color online) Electronic band structures of Na$_2$IrO$_3$ within (a) LDA,
    (b) LDA+SO, and (c) LDA+$U$+SO schemes. Green, red, and blue colored energy dispersions
    in (a) are indicating $e_g$, $e_g'$, and $a_{1g}$ bands respectively, induced by
    the largest cubic and the next largest trigonal crystal fields.} \label{fig:1}
\end{figure}

We investigated the electronic and magnetic structures of Na$_2$IrO$_3$ by
performing LDA, LDA+SO, and LDA+$U$+SO calculations. Calculated electronic
band structure near $E_{\mathrm{F}}$ are shown in Fig.~\ref{fig:1}. The
LDA band structure in Fig.~\ref{fig:1}(a) features the Ir 5$d$ bands of
$e_{g}$ and $t_{2g}$ components separated by a large cubic crystal field
$\Delta_{\textrm{cubic}}\sim$ 4 eV. While narrow $e_{g}$ bands are located
at 3 eV above $E_{\mathrm{F}}$, the top of $t_{2g}$ bands are pinned at
$E_{\mathrm{F}}$ and spread out to -2.0 eV below $E_{\mathrm{F}}$.  Due to
the extended nature of Ir 5$d$ orbitals, there are large contributions to
the band structure from both the indirect hopping via the Ir 5$d$-O 2$p$
hybridization and the direct hopping between the neighboring Ir 5$d$
orbitals.  From the tight-binding analysis \cite{Choong}, even the
next-nearest-neighbor hopping terms through oxygen and sodium atoms make
significant contributions to the LDA band structure.

The trigonal crystal field ($\Delta_{\textrm{trigonal}}$) splits the
$t_{2g}$ bands into $a_{1g}$ and $e_{g}'$ states. In addition, there is a
strong hybridization between neighboring Ir 5$d$ orbitals which gives rise
to the bonding and anti-bonding of $e_{g}'$ orbitals. The bonding and
anti-bonding doublet states consist of $e_{g}'$ orbital pairs of two Ir
atoms per unit cell.  At the $\Gamma$ point of the LDA band structure, the
$e_{g}'$ anti-bonding states, to be called by $e_{AB}$,
are close to $E_{\mathrm{F}}$ while the $e_{g}'$ bonding
states, to be called by $e_{B}$, are at about $-$0.8 eV. The $a_{1g}$
bands located at $-$1 eV have a negligible effect of the hybridization
between neighboring Ir atoms but show a relatively large $c$-axis
dispersion, which may be derived from the character of $a_{1g}$ orbitals
pointing toward the Na atoms in the next layers.  Here it is noted that
the appearance of the $e_{AB}$ doublet at $E_{\mathrm{F}}$ in the LDA band
structure of Na$_{2}$IrO$_{3}$ is in contrast to the presence of almost
degenerate $t_{2g}$ state in Sr$_{2}$IrO$_{4}$ which serves as a basis for
the $j_{\mathrm{eff}}$=1/2 state when SOC is introduced \cite{kim:076402}.

In the LDA band structure, the doubly degenerate $e_{AB}$ states form a
narrow band and cross $E_{\mathrm{F}}$.
The introduction of SOC breaks the degeneracy of $e_{AB}$ by preserving
the time-reversal symmetry so that the $e_{AB}$ bands split off over the
whole Brillouin zone (BZ) as shown in Fig.~\ref{fig:1}(b). Despite the
split of $e_{AB}$ bands, the LDA+SO band structure is still metallic with
a small electron pocket at the $A$ point and hole pockets off the
$k_{c}=0$ plane near $M$. From the tight-binding analysis of the
Na$_{2}$IrO$_{3}$ band structure \cite{Choong}, we obtained
$\Delta_{\textrm{trigonal}}\sim$ 0.6 eV, which is larger than the SOC
parameter $\lambda_{\textrm{SO}}\sim$ 0.4 eV
\cite{kim:076402,PhysRevB.13.2433}. Thus the band structure of
Na$_{2}$IrO$_{3}$ near $E_{\mathrm{F}}$ is characterized by the bonding
$e_{B}$ and anti-bonding $e_{AB}$ states with $\Delta_{\textrm{cubic}}>
\Delta_{\textrm{trigonal}}>\lambda_{\textrm{SO}}$.  Since
$\Delta_{\textrm{trigonal}} > \lambda_{\textrm{SO}}$, however, the
$e_{AB}$ character of the bands are maintained. Contrary to the layered
perovskite Sr$_2$IrO$_4$ system, where the SOC entangles almost degenerate
$t_{2g}$ orbitals with spin states and produces the spin-orbit integrated
$j_{\mathrm{eff}}$=1/2, the strong trigonal field in Na$_{2}$IrO$_{3}$
suppresses the mixing of $a_{1g}$ and $e_{g}'$ states. Instead, the SOC
acting on the $e_{g}'$ subspace plays a role of effective Zeeman coupling,
the details of which will be discussed later. The presence of the
effective Zeeman coupling is manifested in the parallel splitting of
$e_{AB}$ and $e_{B}$ bands.

Similarly to the case of Sr$_{2}$IrO$_{4}$, both the on-site Coulomb
interaction and the SOC are expected to be important in the description of
Ir 5$d$ states. The LDA+$U$+SO band structure shown in Fig.~\ref{fig:1}(c)
was calculated with an effective $U=2.0$ eV, which was found to
be consistent with angle-resolved photoemission and optical spectroscopy
experiments \cite{kim:076402}. As a result of the combined action of both
on-site Coulomb interaction and SOC, a small band gap arises between the
SO-split $e_{AB}$ bands. Two $e_{AB}$ bands form valence and conduction
bands with nearly the same dispersion above and below $E_{\mathrm{F}}$,
respectively. Contrary to the non-magnetic metallic solution of the LDA
and LDA+SO calculations, the LDA+$U$+SO solution predicts an AFM ordering
with local magnetic moments lying within the $ab$ plane. The magnitude of
total moment is 0.47 $\mu_B$ per each Ir atom, which is decomposed into
the spin moment of 0.12 $\mu_B$ and the orbital moment of 0.35 $\mu_B$.

\begin{figure}
 \centering\includegraphics[width=8cm]{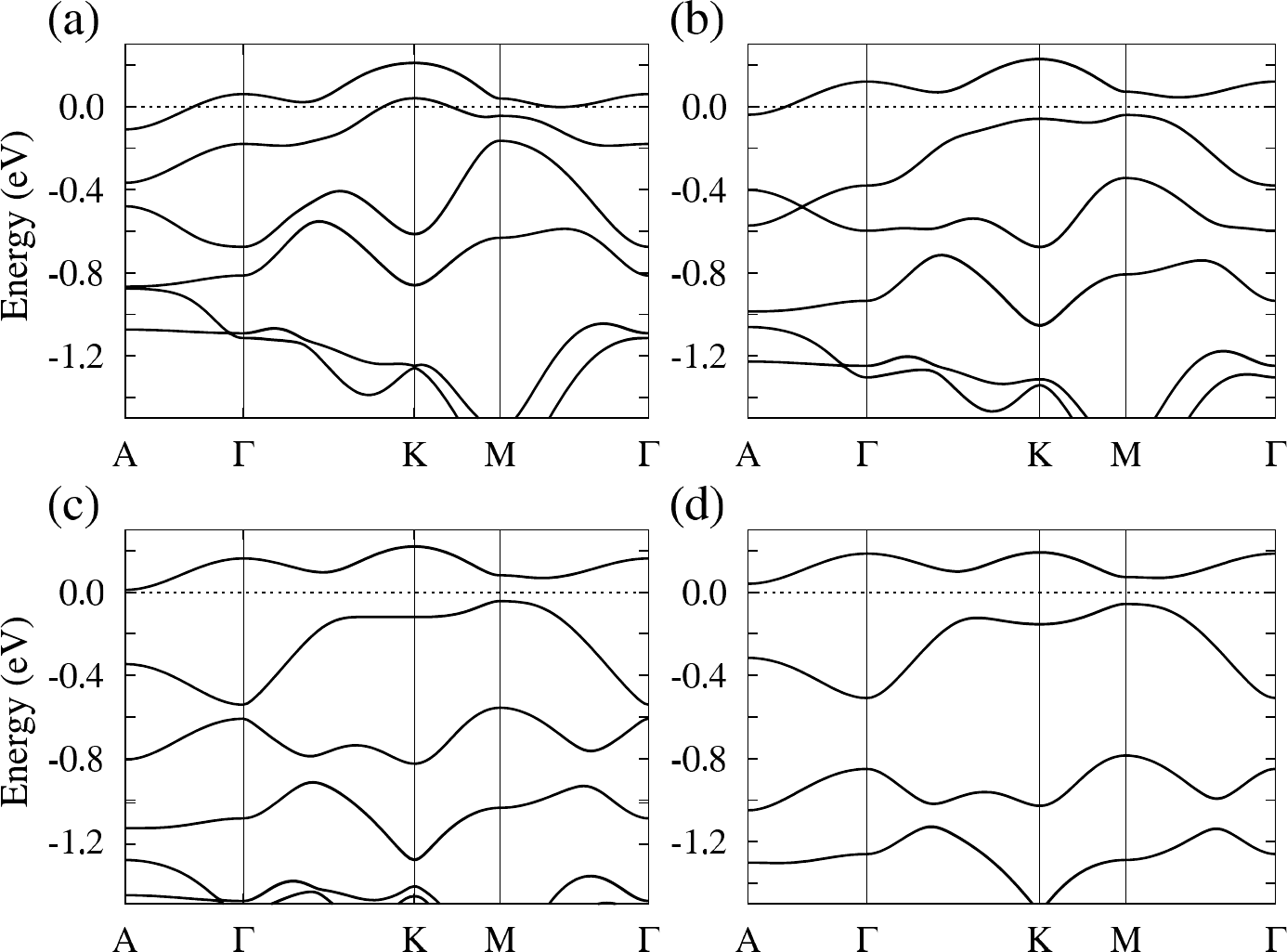}
 \caption{Electronic band structures from LDA+SO calculations with the scaling
    factors of SOC strength $\lambda_{\textrm{SO}}/\lambda_0$ are (a) 0.5, (b) 1.0,
    (c) 1.5, and (d) 2.0, where $\lambda_0$ is the SOI magnitude of a real Ir atom.
    Gap opens when $\lambda_{\textrm{SO}}/\lambda_0$ is increasing from 1.0 to 1.5.} \label{fig:2}
\end{figure}

Despite that the importance of both $U$ and $\lambda_{\mathrm{SO}}$, the
nature of the insulating ground state of Na$_{2}$IrO$_{3}$ is quite
distinct from that of Sr$_{2}$IrO$_{4}$. In Sr$_{2}$IrO$_{4}$, the
$j_{\mathrm{eff}}$=1/2 degeneracy can not be lifted by the SOC and the
Mott-Hubbard gap can be attained only when the on-site $U$ is
introduced. Thus breaking the time-reversal symmetry is essential to get
the insulating ground state of Sr$_{2}$IrO$_{4}$. In the case of
Na$_{2}$IrO$_{3}$, however, the broken time-reversal symmetry is not
required to acquire the insulating state. As shown in Fig.~\ref{fig:1}(b),
the SO-split $e_{AB}$ bands are separated over the whole BZ so that the
increase of the SOC strength can enlarge the already present gap between
two $e_{AB}$ bands. To probe this idea, we carried out DFT calculations by
controlling the SOC strength, which can be achieved by changing the
scaling factor when generating the $j$-dependent pseudo-potential
\cite{openmx}. Calculated results for the scale factors
$\lambda_{\mathrm{SO}}/\lambda_{0}$= 0.5, 1.0, 1.5, and 2.0 are shown in
Fig.~\ref{fig:2}. Taking the original SOC in the real Ir atom as
$\lambda_{0}$ as a reference, $\lambda_{\mathrm{SO}}/\lambda_{0}$= 1.5 was
found to be enough to open a full insulating gap. We call these insulating
ground states as spin-orbit (SO) insulators, which have energy gaps
generated by the SOC. SO insulators have no local moment and preserve the
time-reversal symmetry and thus are distinct from the Mott-Hubbard
insulator.

To understand the origin of SO insulators, we consider the SOC matrix
elements within the $e_g'$ subspace. Since the degenerate $e_g'$ states
can be written by
\begin{eqnarray}
\mid e_1'\rangle=\frac{1}{\sqrt3}(\mid d_{xy}\rangle+e^{\imath\theta}\mid d_{yz}\rangle
+e^{-\imath\theta}\mid d_{zx}\rangle) \nonumber \\
\mid e_2'\rangle=\frac{1}{\sqrt3}(\mid d_{xy}\rangle+e^{-\imath\theta}\mid d_{yz}\rangle
+e^{\imath\theta}\mid d_{zx}\rangle)
\end{eqnarray}
where $\theta=2\pi/3$, the on-site SOC term becomes
\begin{equation}
\langle \mathcal{H}_{\textrm{SO}}\rangle_{e_g'}=
\langle \lambda_{\textrm{SO}}\mathbf{L}\cdot\mathbf{S}\rangle_{e_g'}=\frac{\lambda_{\textrm{SO}}}{2}
\left(\begin{array}{c|c}
\hat{n}\cdot \vec{\sigma} &  \\ \hline
 & \; -\hat{n}\cdot \vec{\sigma}
\end{array}\right)
\end{equation}
where the basis sets are $\mid e_g'\rangle\otimes\mid
S=\frac{1}{2}\rangle=\{ \mid e_1'\alpha\rangle ,\mid e_1'\beta\rangle,\mid
e_2'\alpha\rangle,\mid e_2'\beta\rangle\}$ and $\hat{n}$ is the unit
vector along the $c$-axis, i.e., the [111] direction in the local coordinate
of IrO$_6$ octahedron.  This block-diagonal form comes from the fact that
$\langle \mathbf{L} \rangle$ is simultaneously diagonalized within $e_g'$
manifold and its eigenvalues are $\hat{n}$ and $-\hat{n}$, respectively.
Here the SOC terms in $e_g'$ act as an internal magnetic field
perpendicular to the $ab$-plane. The internal field gives rise to an
effective Zeeman splitting, but the field direction in the $e_{1}'$
component is opposite to that in the $e_{2}'$ component. Thus, the
effective Zeeman coupling does not break the time-reversal symmetry and
$|e_1'\alpha\rangle$--$|e_2'\beta\rangle$ and
$|e_1'\beta\rangle$--$|e_2'\alpha\rangle$ remain as time-reversal
partners. Since the $e_{AB}$ states are the anti-bonding
combination of the $e_{g}'$ orbitals of neighboring Ir atoms, the
splitting of $e_{AB}$ bands by the effective Zeeman coupling
is proportional to the SOC strength as shown in Fig.~\ref{fig:2},
especially at the $\Gamma$ point.

The ground states of Na$_{2}$IrO$_{3}$ with the large SOC strength are SO
insulators. The band gaps are induced by the effective Zeeman coupling of
the SOC within the $e_{g}'$ subspace. Their characters are different from
other types of band insulators such as covalent or ionic ones. The Fermi
level is placed between bonding and anti-bonding bands in covalent
solids and between different ionic configurations in ionic solids. In SO
insulators, the gap is not driven by bonding characters, but mainly
related to the symmetry of the states at $E_{\mathrm{F}}$.  In a sense that
their band gaps are generated by the SOC, SO insulators share the same
ground with topological insulators though it is necessary to prove the
non-trivial topology of its ground state.

One important consequence of
the SO insulating phase is the proximity of the AFM ground state to the SO
insulator state. In the LDA+$U$+SO calculation, the AFM ordered local
moments are aligned in the $ab$-plane. Due to the huge internal field
along [111] direction, it is hard to break the time-reversal symmetry and
to develop local magnetic moments along that direction. Thus, transverse
magnetic moments which are perpendicular to the internal field can be
easily developed.  Strong magnetic anisotropy originated from the internal
magnetic fields might be seen in magnetic susceptibility measurements.

\begin{figure}
 \centering\includegraphics[width=8cm]{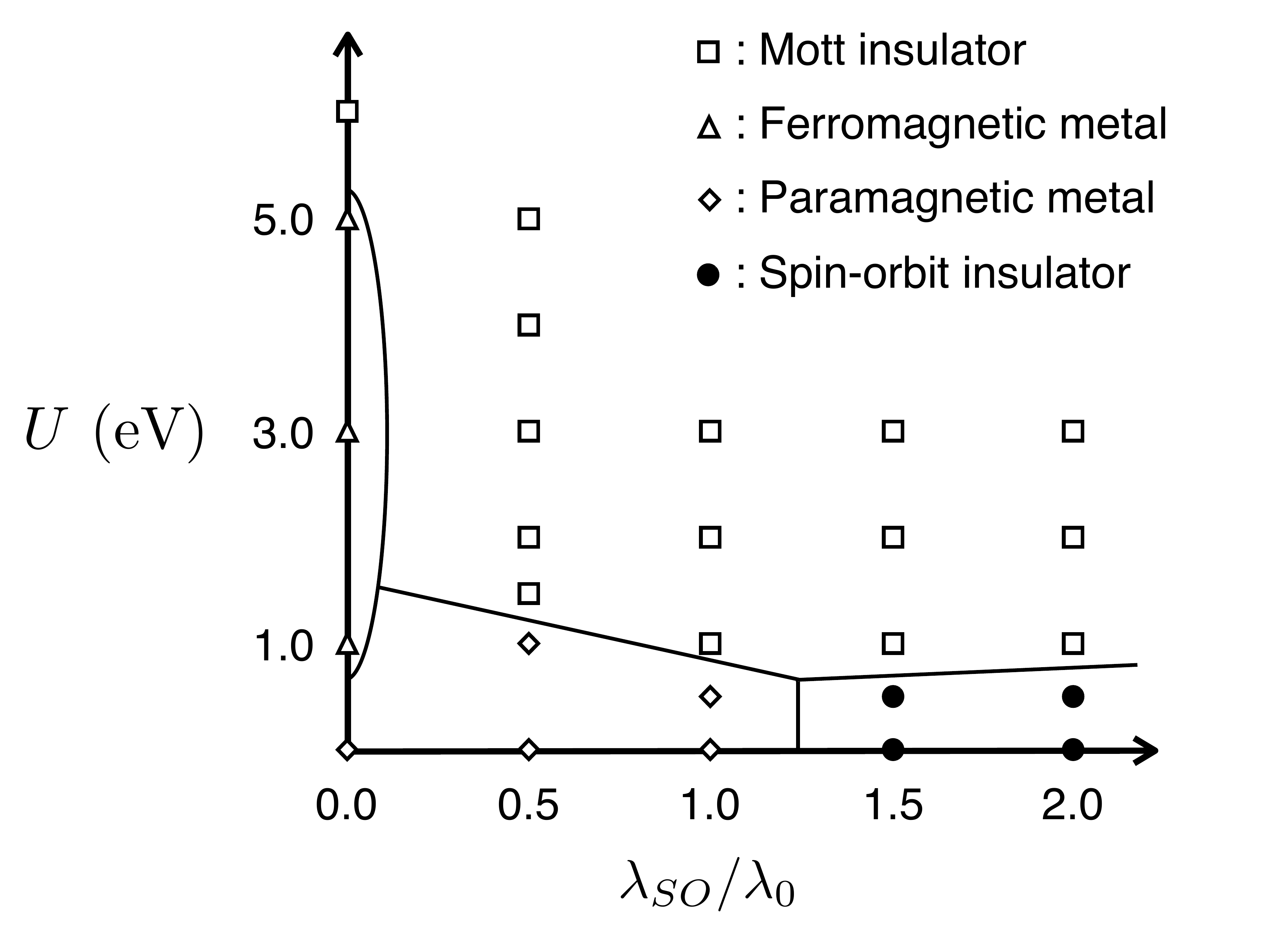}
 \caption{Phase diagram in the $\lambda_{\textrm{SO}}$--$U$ parameter
   space depicting four different phases from LDA+$U$+SO calculations with
   varying $U$ and $\lambda_{\textrm{SO}}$ values. Paramagnetic metallic
   phase appears near the origin, Mott insulator in the region of $U > 1$,
   and SO insulator in the region of $\lambda_{\textrm{SO}}>1$ and $U <
   1$. The real ground state is located inside Mott insulating
   territory.} \label{fig:3}
\end{figure}

To elucidate the relation between SO insulator and AFM Mott insulator
phases, we explored a possible phase diagram of Na$_{2}$IrO$_{3}$ in an
extended $\lambda_{\textrm{SO}}$--$U$ parameter space and present the
result in Fig.~\ref{fig:3}. When $U$ is small and
$\lambda_{\textrm{SO}}/\lambda_{0}$ is less than 1.5, the ground state
remains as a paramagnetic metal. When there is no SOC, i.e.,
$\lambda_{\mathrm{SO}} = 0$, a ferromagentic metallic phase develops in a
narrow range of the parameter space with $\lambda_{\mathrm{SO}} = 0$ upto
$U=5.0$ eV. This ferromagnetic state becomes unstable in the presence of
the SOC.  On the other hand, for the value of $U$ smaller than about 1 eV,
the SO insulator phase emerges as a non-magnetic insulator. Since the band
gap is induced by the effective Zeeman coupling of the SOC within the
$e_{g}'$ subspace, the Kramers degeneracy of the valence states holds up
and the time-reversal symmetry remains unbroken.  For the finite
$\lambda_{\mathrm{SO}}$, Mott insulating AFM states develop as $U$ becomes
larger than about 1.0 eV. The difference between two insulating phases,
i.e., the criterion for the boundary is the existence of local magnetic
moments. The Mott insulating phase has AFM ordering where on-site Coulomb
repulsion breaks the symmetry developing local moments during the
correlation gap opens. Our LDA+$U$+SO calculation predicts that the real
ground state of Na$_2$IrO$_3$ is located in the Mott AFM region with
$U=2.0 \sim 3.0$ eV and $\lambda_{\textrm{SO}}/\lambda_0=1$. However, the
strongly anisotropic nature of its AFM ordering originates from
its proximity to the SO insulator phase.

\begin{figure}
 \centering\includegraphics[width=8cm]{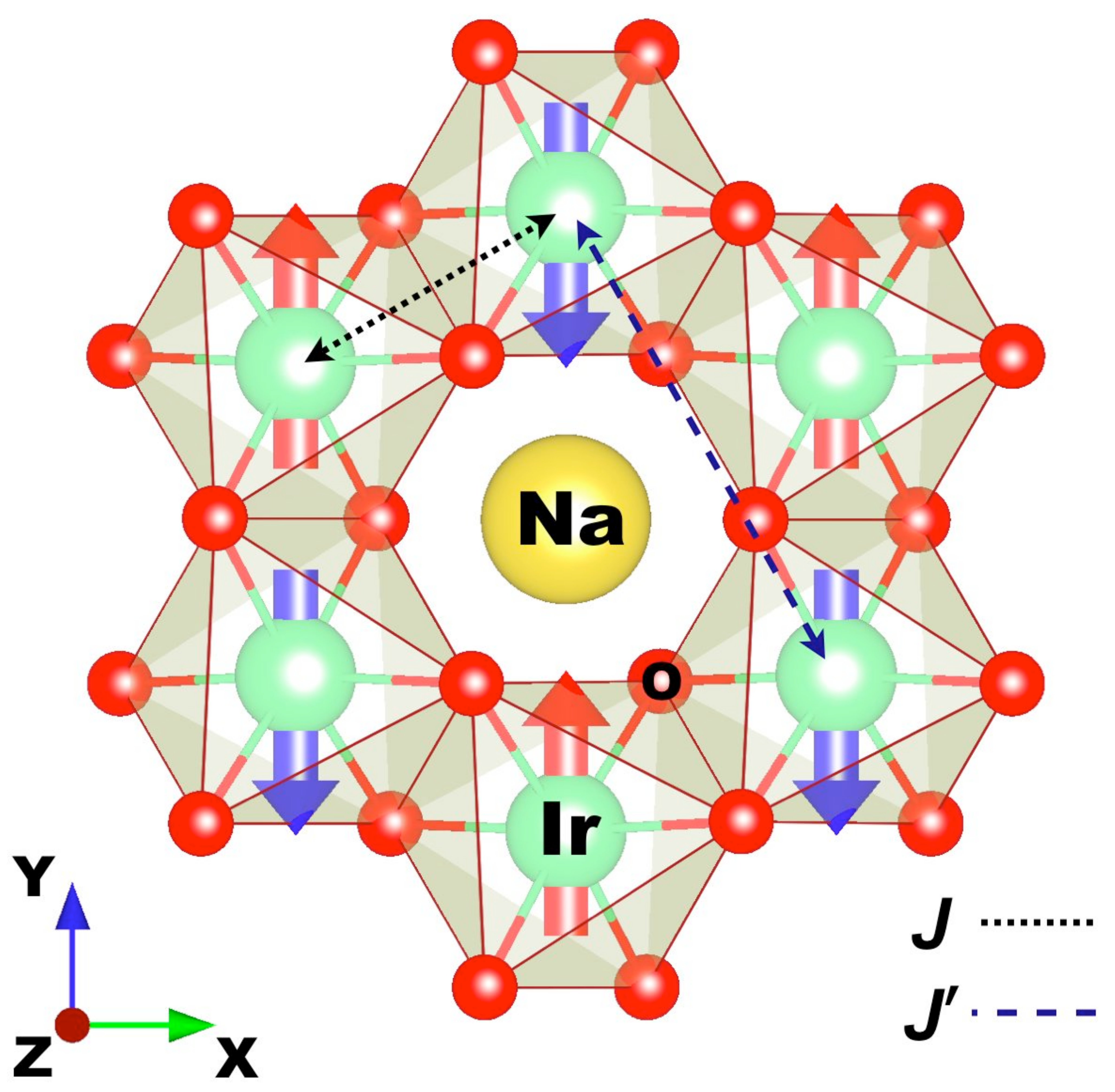}
 \caption{(Color online) Schematic drawing of the
   (Ir$_{2/3}$Na$_{1/3}$)O$_2$ plane and magnetic configuration of the AFM
   insulating ground state of Na$_2$IrO$_3$.  Magnetic moments are ordered
   anti-ferromagnetically lying on the $ab$-plane due to the strong
   internal field along the $c$-axis. Not only the NN exchange $J$ (dotted
   arrow) but the NNN exchange $J'$ (dashed arrow) are significant and may
   give rise to magnetic frustration.} \label{fig:4}
\end{figure}

Another important aspect in Na$_2$IrO$_3$ is magnetic frustration indicated
in large $\theta_{\textrm{CW}}/T_{\textrm{N}}$ ratio from susceptibility
measurements \cite{Takagi}. To reveal the origin of frustration, we have
estimated exchange interactions $J$ and $J'$ between nearest-neighbor (NN)
and next-nearest-neighbor (NNN) Ir atoms respectively.(Fig.~\ref{fig:4})
Calculation scheme is based on the perturbation formalism.
\begin{equation}
J_{ij}=\frac{1}{2\pi}\int^{\epsilon_{F}}d\epsilon \left[
\hat{G}^{\uparrow}_{ij} \hat{V}_j \hat{G}^{\downarrow}_{ji} \hat{V}_i \right],
\end{equation}
where $\hat{G}$ is the one-particle Green's function and $\hat{V}$ is
on-site exchange interaction potential \cite{PhysRevB.70.184421}. The
result is $J'/J=0.47$, which means that NN and NNN exchange coupling
strength are comparable and they might be a source of frustration.  Above
result is mainly attributed to the extended nature of Ir 5$d$ orbitals.
Large direct overlap between NN Ir atoms gives FM direct exchange
interaction, competing with AFM superexchange from oxygen mediated hopping
channels and finally reducing AFM exchange $J$. On the other hand, the NNN
hopping integrals are not negligible that the NNN AFM interaction $J'$ can
be comparable and frustrate long range AFM ordering.

In conclusion we have shown that the spin-orbit entangled $e_g'$ states
under the strong internal Zeeman field driven by the SOC lead to an
unusual band gap. The predicted AFM ground state is in close proximity to
the SO insulator phase where the AFM ordering in Na$_2$IrO$_3$ becomes
strongly anisotropic with quenched moments along the $c$-axis. The highly
anisotropic AFM state in Na$_2$IrO$_3$ may serve as a model system for
the two-dimensional XY model with frustrated exchange interactions.
One may be able to drive a crossover between AFM and SO insulators
through the modulation of structural parameters or chemical substitution,
though we need more study on the role of SOC in the Mott AFM phase in
connection with the topological nature of SO insulators.

\begin{acknowledgments}
  We are grateful to H. Takagi for sharing information prior to
  publication.  This work was supported by the KOSEF through the ARP
  (R17-2008-033-01000-0).  We also acknowledge the support from KISTI
  under the Supercomputing Application Support Program.
\end{acknowledgments}


\end{document}